\documentstyle[11pt,aas2pp4]{article}
\def\Msun{~M_\odot}

\def\kms{\rm ~km~s^{-1}}
\def\ergs{\rm ~erg~s^{-1}}
\def\ml{~\Msun ~\rm yr^{-1}}

\slugcomment{Submitted to the Astronomical Journal}

\lefthead{Fesen et al.}

\begin{document}
\title{Late-Time Optical and UV Spectra of SN~1979C and SN~1980K }

\author{Robert A. Fesen\altaffilmark{1}, Christopher L. Gerardy\altaffilmark{1},
        Alexei V. Filippenko\altaffilmark{2}, Thomas Matheson\altaffilmark{2}, 
        Roger A. Chevalier\altaffilmark{3},
        Robert P. Kirshner\altaffilmark{4}, Brian P. Schmidt\altaffilmark{5}, 
        Peter Challis\altaffilmark{4},
        Claes Fransson\altaffilmark{6}, Bruno Leibundgut\altaffilmark{7}, and
        Schuyler D. Van Dyk\altaffilmark{8}
        }

\altaffiltext{1}{6127 Wilder Laboratory, Physics \& Astronomy Department,
       Dartmouth College, Hanover, NH 03755}
\altaffiltext{2}{ Department of Astronomy,
       University of California, Berkeley, CA 94720-3411}
\altaffiltext{3}{Department of Astronomy, University of Virginia,
       P.O. Box 3818, Charottesville, VA 22903}
\altaffiltext{4}{Harvard-Smithsonian Center for Astrophysics,
       60 Garden Street, Cambridge, MA 02138}
\altaffiltext{5}{Mt. Stromlo Observatory, The Australian National University,
       Private Bag, ACT 2611 Weston Creek P.O.,
       Australia}
\altaffiltext{6}{Stockholm Observatory, SE-133 36 Saltsjobaden, Sweden}  
\altaffiltext{7}{European Southern Observatory, Karl-Schwarzschild-Strasse 2,
       Garching, Germany}
\altaffiltext{8}{IPAC, Caltech, Mailcode 100-22, Pasadena, CA 91125}

\begin{abstract}

A low-dispersion Keck I spectrum of SN~1980K taken in August 1995 
(t = 14.8 yr after explosion)  
and a November 1997 MDM spectrum (t = 17.0 yr) show broad 
5500 km s$^{-1}$ emission lines of H$\alpha$, [O~I] 6300,6364 \AA, 
and [O~II] 7319,7330 \AA. Weaker but similarly broad lines 
detected include [Fe~II] 7155 \AA, [S~II] 4068,4072 \AA,
and a blend of [Fe~II] lines at 5050--5400 \AA.
The presence of strong [S~II] 4068,4072 \AA \ emission but a 
lack of [S~II] 6716,6731 \AA \ emission suggests   
electron densities of 10$^{5-6}$ cm$^{-3}$.  
From the 1997 spectra, we estimate an H$\alpha$ 
flux of $1.3 \pm 0.2 \times 10^{-15}$ erg cm$^{-2}$ s$^{-1}$  
indicating a 25\% decline from 1987--1992 levels  
during the period 1994 to 1997, possibly related to 
a reported decrease in its nonthermal radio emission. 

A May 1993 MMT spectrum of SN~1979C (t = 14.0 yr) shows a somewhat different
spectrum from that of SN~1980K. Broad, 6000 km s$^{-1}$ emission lines
are also seen but with weaker H$\alpha$, stronger [O~III] 4959,5007 \AA, 
more highly clumped [O~I] and [O~II] line profiles,
no detectable [Fe~II] 7155 \AA \ emission, and a faint but very broad
emission feature near 5750 \AA. A 1997 {\it HST} Faint Object Spectrograph,
near-UV spectrum (2200 -- 4500 \AA) shows strong lines of C~II] 2324,2325 \AA,
[O~II] 2470 \AA, and Mg~II 2796,2803 \AA \ along with weak
[Ne~III] 3969 \AA, [S~II] 4068,4072 \AA \ and [O~III] 4363 \AA \ emissions.
The UV emission lines show a double-peak profile
with the blueward peak substantially stronger than the red, suggesting 
dust extinction within the expanding ejecta (E[B--V] = $0.11 - 0.16$ mag).
The lack of detectable [O~II] 3726,3729 \AA \ emission together with
[O~III] (4959+5007)/4363 $\simeq$ 4 imply electron densities 
10$^{6-7}$ cm$^{-3}$. 

These new SNe II-L spectra show general agreement with the lines
expected in a circumstellar interaction model, but the specific models that
are available show several differences with the observations.
High electron densities (10$^{5-7}$ cm$^{-3}$) result in stronger 
collisional de-excitation than assumed in the models, thereby  
explaining the absence of several moderate to strong predicted lines  
such as [O~II] 3726,3729 \AA, [N~II] 6548,6583 \AA, and [S~II] 6716,6731 \AA.
Interaction models are needed that are specifically suited to these
supernovae.
We review the overall observed range of late-time SNe II-L properties and
briefly discuss their properties relative to young, ejecta dominated Galactic 
SNRs.

\end{abstract}

\section{Introduction}

The Type II linear supernova (SN II-L) SN 1980K in NGC 6946 reached a peak 
brightness 
of V = 11.4 in November 1980
(see Barbon, Ciatti, \& Rosino 1982; \cite{HT86} and references therein).
Despite a steadily declining flux through 1982 (\cite{UK86}),
faint H$\alpha$ emission from SN 1980K was detected 
in 1987 through narrow passband imaging (\cite{FB90}, hereafter FB90).
Low-disperson optical spectra obtained in 1988 and 1989 showed  
broad, 6000 km s$^{-1}$ H$\alpha$  
and [O~I] 6300,6364 \AA \ emission, along with 
weaker line emission from [Ca~II] 7291,7324 \AA \ 
and/or [O~II] 7319,7330 \AA,
[O~III] 4959,5007 \AA, and
[Fe~II] 7155 \AA \ (FB90; \cite{U91}; Leibundgut et al. 1991).
Monitoring of its optical flux from 1988 through 1992 indicated a nearly 
constant luminosity (Leibundgut, Kirshner, \& Porter 1993; 
Fesen, \& Matonick 1994, hereafter FM94). 

Following the optical recovery of SN~1980K, a 
handful of other SNe II-L have been optically detected 7 -- 25 yr 
after maximum light. These include SN~1986E in NGC~4302 (\cite{CDT95}),
SN~1979C in M100 (\cite{FM93}), and SN~1970G in M101 (\cite{Fesen93}).
The late-time optical spectra of all these SNe show broad H$\alpha$ and 
forbidden oxygen emission lines, and these characteristics have been 
exploited in searching for fainter, late-time 
SNe II-L (e.g., SN 1985L in NGC~5033; Fesen 1998).

Of the few SNe II-L recovered at late times, SN~1979C in M100 (NGC 4321) 
is particularly noteworthy. This SN was exceptionally bright at outburst for
a Type II event, with a peak $M_{B}$ of --20 mag
(\cite{YB89}; \cite{Gaskell92}), and  became bright in the radio as well, 
eventually reaching a 6 cm luminosity more than 200 times that of Cas A (Weiler 
et al. 1989).
It also exhibits among the highest, late-time H$\alpha$ luminosity
($\sim$ $1 \times$ 10$^{38}$ erg s$^{-1}$)  
and shows radio emission variability which may be due to
a periodic modulation of the progenitor's pre-SN mass loss (\cite{Weiler91}).

Late-time optical SN II emission lines are thought to arise from interactions
with surrounding circumstellar mass loss material (CSM). These interactions
lead to the formation of a reverse shock moving back into the expanding 
ejecta which then subsequently ionizes, either by far-UV or X-ray emission, 
a broad inner ejecta region from 
which the optical lines are produced (Chevalier \& Fransson 1994, hereafter 
CF94).
This model is also consistent with the presence of
accompanying nonthermal radio emission in all optically 
recovered SNe II-L, with radio light curves
like those predicted from the ``mini-shell'' model involving
shock generated synchrotron emission from the forward shock (Chevalier 1982;
Weiler et al. 1993). 

Late-time SN II emission lines are important for the information
they can provide on SNe ejecta abundances, late-time shock 
emission processes, and the mass-loss history 
and evolutionary status of SN progenitors
(Leibundgut et al. 1991; \cite{Weiler91}, Montes et al. 1998).
Unfortunately, optical emission lines from old 
SNe II-L are quite faint ($\leq 3 \times$ 10$^{-17}$ erg 
cm$^{-2}$ s$^{-1}$ \AA$^{-1}$), making accurate measurements 
difficult even using 4-m class telescopes.
Consequently, much is still uncertain about their spectra,
even for SN~1980K, the best and longest studied object.
For example, virtually nothing is known about SN II-L spectra
below 5000 \AA, and uncertainties remain as to whether 
[Ca~II] or [O~II] is chiefly responsible for the strong emission 
commonly seen near 7300 \AA. 
SN~1980K's current optical properties are of special
interest due to a recently reported sharp decline in 
nonthermal radio flux in the interval 1994--1997 (Montes et al. 1998). 

Here we present a 1995 Keck spectrum of SN 1980K which shows  
the region below 5000 \AA \ for the first time, 
and a November 1997 MDM spectrum useful for investigating its recent
H$\alpha$ flux. 
We then compare these SN~1980K data to a 1993 MMT spectrum of SN~1979C at a 
similar age to
the Keck SN~1980K spectrum. We also present a 1997 {\it HST} UV spectrum
of SN~1979C which reveals several strong UV and near-UV lines
which help clarify the nature of the observed emission.
Finally, we outline some general observed properties of late-time SNe II-L
and compare them with young Galactic supernova remnants (SNRs).
 
\section{Observations}

  Two consecutive low-dispersion spectra of SN 1980K were obtained on 28
November 1995 with the Low Resolution Imaging Spectrometer (LRIS; \cite{Oke95})
at the Cassegrain focus of the Keck-I telescope. The duration of each
exposure was 30 minutes. Conditions were not photometric, and the seeing was
about 1$''$. The position angle of the slit (of width $1''$) was 128.5 degrees,
through two stars that are known to be colinear with the supernova position;
this was also close to the parallactic angle at the time of observation. We
used a Tektronix 2048 x 2048 pixel CCD with a scale of $0.41''$ per binned pixel
in the spatial direction, a gain of 1.1 e$^{-}$/count, and a readout noise of 6
e$^{-1}$/pixel.

  Cosmic rays were eliminated from the two-dimensional spectra through a
comparison of the pair of exposures. The background sky was measured in regions
adjacent to the supernova, and subtracted from the extracted spectrum.  Spectra
of Hg-Kr-Ne-Ar comparison lamps were used to determine the wavelength scale and
resolution ($\sim 10$~\AA). Flux calibration and removal of telluric absorption
lines were achieved with a spectrum of the sdF star BD+174708 (\cite{OG83}.
 
A spectrum of SN~1980K was also obtained on 3 November 1997
using the 2.4 m Hiltner telescope at the MDM Observatory on Kitt Peak.
A Modular Spectrograph like that in use at
Las Campanas Observatory was employed with a N-S aligned $2.2'' \times 4'$
slit, a 600 line mm$^{-1}$ 5000 \AA \ blaze grating, and a 1024 x 1024
Tektronics CCD detector. A single 6000 s exposure covering the spectral region
4000 -- 8500 \AA \ was taken with a spectral resolution of about 5 \AA.
Cosmic rays were removed using standard IRAF routines for pixel rejection
and replacement. Observing conditions were photometric but due to 
slit light losses caused by variable seeing during
the long exposure, absolute fluxes are reliable only to $\pm15$\%.

A spectrum of SN~1979C was obtained on 21 May 1993 using the
Red Channel long-slit CCD spectrograph (Schmidt, Weymann, \& Foltz 1989) on the
4.5-m Multiple Mirror Telescope (MMT).
Three 1200 s exposures were taken using a $2'' \times 180''$ slit 
and covering a spectral range 3800 -- 9900 \AA \ with 12 \AA \ resolution.
A strong blue continuum, possibly from O and B stars near SN~1979C's location
in NGC 4321 (M100), was removed in the final reduction and an 
arbitrary zero flux level set. Because of the greatly increased noise level 
redward of 8000 \AA, we show here only the region 4000 to 8000 \AA.

An {\it HST} spectrum of SN 1979C was obtained using the Faint Object
Spectrograph (FOS) on 30 January 1997. 
Three G270H exposures (2200 -- 3275 \AA) were taken for a total time
of 7170 s, plus one G400H spectrum (3250 -- 4750 \AA) with an
exposure of 2410 s.
Both the G270H and G400H data shown here have been smoothed by a 5 point 
average.

\section{Results}

\subsection{SN 1980K}

The Keck and MDM spectra of SN~1980K are shown in Figures 1 and 2 respectively.
Though the Keck spectrum has significantly higher S/N, both 
show broad H$\alpha$ and [O~I] 6300,6364 \AA \ lines 
plus emission near 7100 \AA \ and 7300 \AA.
These features have been seen in earlier SN~1980K spectra
with roughly the same relative strengths and widths as seen here.
However, the line profiles are better defined in the Keck data and the spectrum 
reveals other fainter emission lines not previously seen.  
 
{\bf H$\alpha$:} SN 1980K's late-time optical flux was first detected 
in 1987 via its strong H$\alpha$ emission and it 
remains among the strongest optical lines observed a decade later.
The 1995 Keck spectrum shows the H$\alpha$ line profile
with an expansion velocity of --5700 to +5500 km s$^{-1}$ with
a strong asymmetry toward the blue, in good agreement with
earlier 1992 and 1994 measurements (FM94; Fesen, Hurford, \& Matonick 1995, 
hereafter FHM94).
A comparison of SN~1980K's H$\alpha$ line profile and strength changes
over the last ten years is shown in Figure 3 using the Lick
June 1988 spectrum (FB90) and our November 1997 MDM data.
Besides illustrating a drop in H$\alpha$ flux over this time period
(see Section 4 below),
one also sees a substantial decrease in the line width, most noticeable
toward the blue edge. In 1988, the FWHM of H$\alpha$ was around  
220 \AA, compared to 190 \AA \ in 1997.
This decrease is consistent with earlier
measurements (FM94) and is predicted by SN -- CSM interaction models (CF94). 
However, the asymmetric profile of H$\alpha$ appears not to have changed
much during the last ten years, with a sharp blue emission edge and an emission 
peak
near 6500 \AA \ visible in both the 1988 and 1997 line profiles.

A faint, unresolved line is seen on top of 
the broader H$\alpha$ emission around zero radial velocity.
A narrow H$\alpha$ emission feature was evident in the 1988 Lick spectrum,
but much less so in subsequent data (FM94; FHM95).
Well detected again here, this narrow H$\alpha$ emission 
may be due to a small H~II region near the SN site
or ionized wind material associated with the progenitor (FB90; FHM95).

From the 1997 MDM spectrum, we estimate a broad emission H$\alpha$ flux of
$1.3 \pm 0.2 \times 10^{-15}$ erg cm$^{-2}$ s$^{-1}$.
This is close to the 1.4 $\times$ 10$^{-15}$ erg cm$^{-2}$ s$^{-1}$ value 
reported by FHM95 for a 1994 spectrum and supports the
conclusion of a $\simeq$ 25\% fading from
flux levels observed in the late 1980s and early 1990s.
 
{\bf H$\beta$:} There is broad emission near 4850 \AA, some of
which may be due to H$\beta$ 4861 \AA.  
Its line strength, however, is difficult to disentangle from other 
possible emission lines between 4800 and 5000 \AA \ (see below). 
Nonetheless, the observed H$\alpha$/H$\beta$ line ratio must be $\geq$ 6, 
suggesting, after correcting for a foreground reddening 
of E(B--V) = 0.40 mag (\cite{BH82}), an intrinsic value $\geq$ 4.

{\bf H$\gamma$:} Weak emission near 4350 \AA \ may be attributable
in part to H$\gamma$ emission, blended with
[O~III] 4363 \AA.

{\bf [O~I]:} The broad [O~I] 6300,6364 \AA \ emission  
is asymmetric towards the blue with a steep blue edge and   
a more gradual decline toward the red. The line's  full velocity 
range is --6000 km s$^{-1}$ if attributed to 6300 \AA \ on the blue side,
and $\geq$ +1700 km s$^{-1}$ if attributed to 6363 \AA \ on the red. 
The blueward velocity is larger than the --4800  km s$^{-1}$ 
estimated by FM94, probably due in part to weaker emission 
detected due to the higher S/N in the Keck spectrum.
In addition, a prominent emission peak at 6280 \AA \ 
suggests strong clumping of 
the O-emitting material in the SN's facing expanding hemisphere.

{\bf [O II]:} One of the strongest emission features in the spectrum, rivaling 
H$\alpha$ and [O~I], lies near 7300 \AA, and extends from 
7180 to 7420 \AA. It has been alternatively identified
as [Ca II] 7291,7324 \AA \ (\cite{U91}; FHM95), [O~II] 7319,7330 \AA \
(in SN1979C; FM93), or a combination of the two (FB90; FM94). 
CF94 suggest it is likely to be solely [Ca II] 
since these [O~II] lines are weak in a variety
of SN--CSM models and in a variety of 
photoionized objects. However, these models also predict
appreciable Ca~II infrared triplet emission (8498, 8542, 8662 \AA)
which is absent in SN~1980K's near-IR spectrum (FM94; FHM95).

Figure 4a shows SN~1980K's [O~I] 6300,6364 \AA \ line profile in terms of
expansion velocity relative to 6300 \AA \ compared 
to the 7300 \AA \ feature if identified as  
the [O~II] line blend at 7325 \AA \ (top panel),
or [Ca~II] using the stronger 7291 \AA \ [Ca~II] line (bottom panel).
The [O~II] blend interpretation is seen to provide a better match
to [O~I]'s line width and its emission substructure than a [Ca~II] 
interpretation. 
For instance, the 6280 \AA \ emission peak in the [O~I] line profile 
matches well with the 7300 \AA \ feature's 
emission feature at 7300 \AA \ if interpretated as an [O~II] blend.
In addition, the velocity of the 7300 \AA \ feature's FWHM blueward edge agrees
within measurement uncertainties to those of [O~I] and H$\alpha$
using the 7319 \AA \ [O~II] line:
namely, 4750, 4500, and 4600 km s$^{-1}$ for [O~I], H$\alpha$, and 7300 \AA \
respectively, compared to 3450 km s$^{-1}$ using [Ca~II] 7291 \AA.

A possible complication is an atmospheric absorption band at 
7190 \AA \ which might have affected the 7300 \AA \ feature's blue emission. 
This telluric absorption band is not especially strong, however,  
and though it could help to explain the poor [Ca~II] profile match to that of 
[O~I] at large negative velocities (see Fig. 4), there still would remain the 
discrepancy along the line's red end. 
We therefore conclude that [O~II] 7319,7330 \AA \ emission
is the dominant cause for the 7300 \AA \ emission feature.

{\bf [O III]:} 
Previous SN 1980K spectra detected broad emission near 5000 \AA \
with an emission peak near 4955 \AA. The broad emission 
was initially identified as   
[O~III] 4959,5007 \AA \ emission (FB90; \cite{Lei91}; FM94) although  
\cite{U91} suggested it might be a blend of permitted Fe~II lines.
As seen in Figure 1, broad emission extends from 
4750 to 5020 and from 5100 to 5400 \AA, with 
little emission between 5020 and 5100 \AA. 
Absence of strong emission just redward of 5007 \AA \ is 
hard to reconcile with a broad, 5000 km s$^{-1}$
[O~III] line profile like seen for [O~I], [O~II], and H$\alpha$ unless the 
[O~III] emission is weak and has a strong blue asymmetry.
In fact, the sharp cutoff of emission at 5020 \AA \ places an expansion limit 
on strong [O~III] 5007 \AA \ emission of $\leq +1000$ km s$^{-1}$.
Nonetheless, [O~III] appears a much more likely identification than
either permitted or forbidden Fe~II and Fe~III lines (cf. FHM95),
especially in view of the clear [O~III] emission in other
SNe II-L (e.g., SN 1979C; see Section 3.2).
The far blue side of the feature is probably due to
H$\beta$ 4861 \AA.
Finally, weak emission near 4350 \AA \ may be a blend of
[O~III] 4363 \AA \ and H$\gamma$.

The Keck spectrum also shows two narrow lines at 4957 and 5005 \AA. 
These are probably related to the narrow H$\alpha$ emission observed.
However, in many earlier spectra, only one  
emission spike, around 4950 -- 4960 \AA, could be clearly seen
(FB90; \cite{Lei91}; FM94) and was interpreted as possible evidence 
of emission substructure in the 4959,5007 [O~III] blend.
An alternative explanation is that the narrow 4959 \AA \ line emission
happens to coincide 
with a $-3600$ km s$^{-1}$ emission feature in the broad [O~III] line,
like that seen in the [O~I] and [O~II] line profiles.

{\bf [Fe II]:} Emission near 7100 \AA \ which merges
into the red wing of the 7300 \AA \ line 
is likely [Fe~II] 7155 \AA \ (FB90; FM94).
Adopting a line center of 7100 \AA \ gives an emission centroid velocity
of --2300 km s$^{-1}$. Emission extends toward the blue 
as far as $\simeq$ 7010 \AA \ suggesting a maximum 
velocity of --6000 km s$^{-1}$.
Due to the blend with the [O~II] 7300 \AA \ feature, little can be
determined regarding its maximum recession velocity. However, the line
extends at least to 7175 \AA \ (+800 km s$^{-1}$).

Broad weak emission from 5020 to 5400 \AA \ can 
be understood in terms of blends of several 
[Fe~II] lines -- specifically the 5112 (19F), 5158 (18F), 5159 (19F),
5262 (19F), 5269 (18F), 5273 (18F),
and 5334 (19F) \AA \ lines. These represent all of the stronger [Fe~II] lines
from the multiplets 18F and 19F typically seen in shocked supernova remnant 
emissions (cf. Fesen \& Hurford 1996).  Adopting a line center shift
of --2300 km s$^{-1}$  and a velocity width of
--6000 to +3500 km s$^{-1}$ like that seen for the [Fe~II] 7155 \AA \ line,
a blend of [Fe~II] lines can explain the 5020--5400 \AA \ emission: 
specifically, its width, 
peak intensity near the strong 5158 and 5159 [Fe~II] lines, and the 
lack of emission longward of 5400 \AA \ and shortward of 5050 \AA.

{\bf [S II]:} The weak emission feature near the blue end of the Keck spectrum 
appears to be [S~II] 4069,4076 \AA. Adopting this identification,  
sulfur has a velocity range of $-5500$ to +4000 km s$^{-1}$, comparable
to the broad H and O lines. As with other lines, the [S~II]
profile also shows a strong blue asymmetry, with a line center shifted to the 
blue
at least --2000 km s$^{-1}$. Finally, we note there is no obvious emission 
from the nebular [S~II] 6716,6731 \AA \
lines in either the Keck or MDM spectra, and previous SN~1980K 
studies have failed to detect any
[S~III] 9069,9531 \AA \ emission (FHM95).

\subsection{SN 1979C}

\subsubsection{Optical Spectrum}

A 1993 MMT spectrum of SN~1979C  
taken 14 yr after maximum light, or roughly the same age 
as SN~1980K when the 1995 Keck spectrum was obtained,
is shown in Figure 5. The spectrum is similar to those presented by FM93 
but with much improved S/N especially in the blue.
 
{\bf H$\alpha$:} Broad H$\alpha$ emission is detected with an 
expansion velocity of $\pm 6200$ km s$^{-1}$.
A strong, narrow  H$\alpha$ component is also seen at +120 velocity in M100's
rest frame (V = +1570 km s$^{-1}$). 
The broad H$\alpha$ emission has a flux of $3 \pm 0.5 \times 10^{-15}$ erg 
cm$^{-2}$ s$^{-1}$,
or slightly larger than the 
$2.5 \times 10^{-15}$ erg cm$^{-2}$ s$^{-1}$
reported by FM93. 
H$\alpha$'s  peak flux is somewhat weaker 
than [O~I] 6300 \AA \ and the profile shows a strong blue asymmetry. 
Narrow [S~II] 6716,6731 \AA \ emission is seen along its red edge,
presumably associated with the narrow H$\alpha$ emission.

{\bf [O I]: } SN~1979C's late-time [O I] 6300,6364 \AA \ emission profile is 
double-peaked much like 
that reported by FM93 but with different peak velocities. 
From the MMT spectrum, we measure peaks at 6230 and 6320 \AA \ 
($-4900$ and $-600$ km s$^{-1}$ if attributed to 
just the 6300 \AA \ line), whereas FM93 found peaks at 6214 and 6329 \AA \ 
($-5700$ and $-190$ km s$^{-1}$).
However, [O~I]'s full expansion velocity in the MMT spectrum 
is $-6300$ to $-1300$ km s$^{-1}$
and in good agreement with FM93.
A June 1998 2.4m MDM [O~I] image of SN~1979C taken using the same filter used by
FM93 indicates a 10\% increase in [O~I] luminosity between 1991.5 and 1998.5. 

{\bf [O II]: } FM93 identified the emission near 7300 \AA \ as [O~II]
7319,7330 \AA \ based upon the excellent match of the 
feature's double-peaked velocity profile with the [O~I]
emission peaks. Two strong emission peaks can also be seen in 
the MMT spectrum at 7240 \AA \ and 7347 \AA. These correspond to
$-5000$ and $-700$ km s$^{-1}$ in M100's rest frame, 
in agreement with the above measured [O~I] peak velocities.
This is graphically shown in Figure 4b and helps validate 
the [O~II] identification made for the SN~1980K spectrum.
We note that the [O~II] 7319,7330 \AA \ emission 
appears stronger in 1993 relative to [O~I] when 
compared to 1991/1992 data (FM93),
making it now the strongest feature in SN~1979C's optical spectrum.

{\bf [O III]: } The MMT spectrum of SN~1979C also shows broad, strong [O~III] 
emission
centered at 4965 \AA \ and spanning 4880 to 5060 \AA, which translate to 
$-6300$ to +1600 km s$^{-1}$. No emission peaks like those seen for
[O~I] and [O~II] are evident from these data.
We measure an observed [O~III] flux of 
$3.5 \times 10^{-15}$ erg cm$^{-2}$ s$^{-1}$.

Broad Emission at 5750 \AA: 
Weak but very broad emission is also  
seen near 5750 \AA \ and extending from
5550 to 6000 \AA. Though faint, this emission is consistent 
with an earlier report of broad emission at 5700--5900 \AA \ by FM93.
Possible identifications include blends of [N~II] 5755 \AA,
Na~I 5890,5896 \AA, and He~I 5876 \AA.

\subsubsection{HST FOS Spectrum}

Figure 6 shows the combined 1997 FOS spectra (G270H and G400H)
of SN~1979C covering the wavelength region 2200 to 4750 \AA.
The continuum levels at the G270H/G400H 3250 \AA \ crossover point agreed
quite well and no zero-point correction was applied.

Three prominent emission lines were detected below 3000 \AA: 
C~II] 2324,2325 \AA, 
[O~II] 2470,2470 \AA,
and Mg~II 2796,2803 \AA. 
These lines show clumpy and strongly asymmetric profiles with the blueward
peaks about twice as bright as those along the redward side, suggestive of
internal dust extinction of the receding emission regions. 
Sharp blue emission peaks in these lines match well those seen
in the optical lines from the MMT spectrum; specifically,
the C~II], [O~II] and Mg~II emission peaks at
2297 \AA, 2440 \AA, and 2767 \AA \ correspond to
the $\sim -5000$ km s$^{-1}$ peaks in the 
[O~I] 6300 \AA \ and [O~II] 7325 \AA \  optical lines.
The Mg~II 2800 \AA \ line appears broader than the two other UV lines, and
this might indicate the presence of weak Mg~I 2852 \AA \
along its red wing. 
It is also possible that weak Si~II] 2335 \AA \ emission might also be blended 
with 
the C~II] 2324,2325 \AA \ feature.

The [O~II] 2470 \AA \ lines arise from the
same upper level ($2p^{2}$ $^{2}P$) as the near-IR 
[O~II] 7319,7330 \AA \ lines. The relative strength of the 2470/7325 
ratio is determined by radiative transition probabilities
(0.78; Mendoza 1983) and thus can be used
to derive an estimate of the foreground extinction.
Assuming no contribution from [Ca~II] 7291,7330 \AA,
the observed 2470/7325 ratio of $0.21 \pm 0.02$ suggests 
that E(B--V) = $0.27 \pm 0.2$ mag.
However, this includes dust extinction within the SN's ejecta as
indicated by the weaker red emission peaks in these UV lines
compared to the near infrared lines. Therefore, measuring only the 
blue emission peaks in the [O ~II] 2470 and 7325 \AA \ lines,
we find a 2470/7325 ratio of $0.25 \pm 0.02$ which implies
E(B--V) = $0.23 \pm 0.02$ assuming R = A$_{\rm V}$/E(B--V) = 3.1.
This should be a reasonable estimate of the foreground extinction 
(both Galactic and in M100)
for SN~1979C. However, it is larger than the previous total reddening value
of $0.18 \pm 0.04$ estimated toward SN~1979C (\cite{DV81}) and may indicate
that some weak [Ca~II] 7291,7330 \AA \ emission contributes to the 
7300 \AA \ feature's strength. This would make the 2470/7325 ratio 
appear smaller and hence the reddening larger.
Alternatively, there might have been changes in [O~II]
line strengths during the four year interval between when the optical (1993)
and UV (1997) data were taken.

One can also estimate the internal dust extinction 
in the young, developing remnant by comparing the red emission spikes
in the [O~II] 2470 and 7325 lines. 
The observed 2470/7325 ratio of 0.15 suggests E(B--V) = $0.34 \pm 0.05$
implying an internal reddening of E(B--V) = $0.11 - 0.16$ 
in addition to the foreground E(B--V) of $0.18 - 0.23$.  
Although the evidence for internal extinction is strong due to 
the consistently weaker red emission spikes in the UV line profiles, our
measurement is not very accurate from this single measurement. 
Although the red emission peak is comparable or slightly stronger than the blue
in the optical lines of [O~I] 6300,6364 and [O~II] 7319,7330,
much greater variation in blue/red peak strengths is seen for the UV lines  
in the {\it HST} data.
For example, the red emission peak in the
[O~II] 2470 \AA \ line is about 45\% as strong as the blue
emission peak, around 65\% in the
C~II] 2324 \AA \ line, and less than 35\% for the Mg~II 2800 \AA \ line.

Emission features detected between 3000 and 4600 \AA \ in the short 
G400H exposure are considerably weaker and 
consequently have less certain identifications but appear to be 
[Ne III] 3869 \AA, [S~II] 4068,4076 \AA, and 
[O~III] 4363 \AA. The strength of the temperature sensitive
[O~III] 4363 line is surprisingly large relative to the [O~III]
4959,5007 lines and suggests  
electron densities $>$ 10$^{6}$ cm$^{-3}$. The only other viable line 
identification for the 4335 \AA \ emission is
H$\gamma$. This seems unlikely, however, due to the weakness
of the H$\alpha$ emission.
Finally, our detection of [S II] 4068,4076 \AA \ 
is tentative but would be consistent with the SN~1980K spectrum.  

\section{Discussion}

To date, five papers have been published on the late-time 
optical spectrum of SN~1980K plus one each on SN~1970G, SN~1979C,
and SN~1986E. None of these, however, presented spectra of 
sufficient S/N or covered a wide enough wavelength range to 
constrain several basic physical parameters such 
as densities or internal extinction, or test current late-time emission models.
The new SN~1979C and SN~1980K spectra do provide some of this 
information and give a much clearer picture for 
late-time optical and UV properties of SNe II-L.
In addition, because SN~1979C and SN~1980K were observed at comparable ages, 
they 
can be used to gauge the spread of SN II-L late-time emission properties.

\subsection{Late-Time Optical Properties of SNe II-L}

Table 1 lists the observed emission properties of
four Type II-L SNe having detected late-time optical emission. 
They can be seen to exhibit several spectral similarities.
From $\sim$ 8 yr and extending at least to  
17 yr, the strongest optical emission lines in SNe II-L 
are H$\alpha$, [O~I] 6300,6364, [O~II] 7319,7330, and [O~III] 4959,5007.
These lines show similar expansion velocities,
$\sim$ 5000 -- 6000 km s$^{-1}$, often with roughly equal line strengths.
Several of these lines also have emission profiles showing evidence 
for dust formation by virtue of diminished 
flux from the receding emission portions. 
The presence of this dust extinction creates 
both blueshifted line centers and asymmetric blue line profiles.
In addition, all four SNe show narrow emission lines (e.g., H$\alpha$ and 
[O~III])
which could imply similar interstellar environments for the progenitors such as
local H~II regions.

However, significant spectral differences also exist.
H$\alpha$ luminosities cover more than an order of magnitude,
with a factor of six difference between objects 
of similar age (e.g., 79C and 80K).
SN~1970G and SN~1986E exhibit little if any broad [O~III] emission, SN~1980K 
some,
while in SN~1979C [O~III] is stronger than H$\alpha$. Conversely,
SN~1980K has fairly strong [Fe~II] emission, both at 7155 \AA \ and
the blend at 5100--5300 \AA, whereas SN~1979C and the other two do not. 
In addition, the very broad, unidentified emission near 5750 \AA \ in SN~1979C 
is not seen in SN~1980K.

Bright optical emission appears correlated with the presence of strong late-time 
radio
nonthermal emission and inferred high mass loss rates (see Table 1)
in the sense that H$\alpha$ luminosity tracks fairly well the peak 6 cm radio
luminosity and the derived mass loss rate from late-time radio data. For 
example, 
SN~1979C is the brightest radio and optical SN and has the largest
estimated mass loss rate while SN~1970G is optically the faintest with the 
lowest
estimated mass loss rate.
However, this correlation is not a particularly tight one. 
SN~1986E is nearly as bright optically as SN~1979C, yet it is
quite faint in the radio.
Moreover, strong radio emission is also not always a predictor of bright optical 
emission.
Two of us (Fesen \& Filippenko) have been 
unsuccessful in detecting 
late-time optical emission from SN~1968D in NGC 6946 (the same parent
galaxy as SN~1980K) as a follow-up to its late-time radio emission recovery 
(Hyman et al. 1995). The lack of detectable optical emission may be
related to SN~1968D's location closer to NGC~6946's center and in a dustier
region (see \cite{Trewhella98} and \cite{VanDyk98a}) which may be attenuating 
faint, 
late-time optical emission below the detection threshold.

Late-time, optical emission lines from SNe II-L appear surprisingly 
steady over time spans of many years. Our 1993 spectrum of SN~1979C indicated
no decrease in H$\alpha$ flux compared to 1990/1991 data (FM93) and actually 
showed
a slight increase as does a comparison of 1991/1998 [O~I] images. 
Recent radio measurements indicate
a flattening of its radio light curve (Van Dyk et al. 1998b) and new 
optical measurements should be undertaken. SN~1980K's H$\alpha$ flux 
remained nearly constant from 1987 through 1994, decreasing only recently.
In Figure 7, we  show the H$\alpha$ light curve over one decade:
from 1987 through 1997.
The data indicate no change between 1987 and 1992.5.
Spectra taken in 1991.4 and 1992.5, though indicating possibly 
a slow fading, were of lower quality than
one obtained in 1992.6 which indicated no significant drop in H$\alpha$ strength 
(FM94).
Moreover, broadband photometric studies covering the years 1990 -- 1992 
indicated
SN 1980K's optical flux level remained virtually unchanged from that seen
in 1987, especially in the R band which is sensitive to H$\alpha$ emission
(R = 21.9 $\pm$ 0.1 mag; Leibundgut et al. 1993).
 
A more recent spectrum taken in 1994, however, showed
an H$\alpha$ emission flux of
$(1.4 \pm 0.2) \times 10^{-15}$ erg cm$^{-2}$ s$^{-1}$ (FHM95),
about 25\% less
than the $1.7 \times 10^{-15}$ erg cm$^{-2}$ s$^{-1}$ found earlier.
Our 1997 measurement of $(1.3 \pm 0.2) \times 10^{-15}$ erg cm$^{-2}$ s$^{-1}$
supports a fading H$\alpha$ strength which may have started
during the last several years, possibly in 1994 with the FHM95 reported decline.
While this optical fading would roughly coincide with a drop in radio
emission (Montes et al. 1998), we cannot completely rule out a small,
steady decline in H$\alpha$ over the 1987--1997 period from the spectroscopic 
data.

During t = 10 -- 20 yr, SN~1980K's broad H$\alpha$ line profile gradually 
narrowed,
changing from a FWHM = 220 \AA \ in 1988.6 to 
190 \AA \ in 1997.9 (see Fig. 3).
This trend has been noted earlier (FM94) and is predicted by
SN--CSM interaction models (CF94).
However, H$\alpha$ appears to be the line most affected, with the [O I] 
6300,6364 line
profile changing little during this time interval.

With the notable exception of H$\alpha$, all the broad emission lines
exhibit a spiky profile suggestive of clumpy emission regions at
particular velocities. This is perhaps most clearly seen in SN~1979C where
emission peaks at $-5000$ and $-1000$ km s$^{-1}$ can be seen in
the [O~I] 6300, [O~II] 2470, [O~II] 7325, C~II] 2324, and Mg~II lines.
The less blueshifted emission peaks are always weaker in the UV lines
compared to the red or near infrared lines, consistent with the presence of 
internal
extinction. In a similar fashion, SN~1980K also exhibits strong emission peaks 
in the
[O~I] 6300 and [O~II] 7325 lines. 
The fact that the H$\alpha$ profile in both SN~1979C and SN~1980K does not 
show such emission peaks suggests that
the O, C, Mg-rich emitting material is coming mostly from ejecta and is 
physically separate from the dominant H$\alpha$ emitting material 
which includes the swept-up shell.

Our new spectra also allow one to
get a handle on the electron density and temperature.
In SN~1980K, we detected [S~II] 4068,4076 \AA \ lines but not 
[S~II] 6716,6731 \AA \ emission. The [S~II] 4068,4072 lines
have a critical density, n$_{\rm cr}$, of  $1.3 \times 10^{6}$ cm$^{-3}$, 
whereas
the 6731 \AA \ line has n$_{\rm cr}$ = $1.5 \times 10^{4}$ cm$^{-3}$.
This suggests collision de-excitation of the [S~II] 6716,6731 lines due to  
electron densities near 10$^{5-6}$ cm$^{-3}$. An upper limit of $\sim$ 10$^{7}$ 
cm$^{-3}$ 
is suggested by the presence of strong
[O~I] 6300,6364 \AA \ emission
(n$_{\rm cr}$ = $1.6 \times 10^{6}$  cm$^{-3}$)
and [O~II] 7319,7330 \AA \
(n$_{\rm cr}$ = $3 \times 10^{6}$ cm$^{-3}$), although [O~I] 
6300,6364 \AA \ emission can be strong even at high densities because it
is an important coolant for low ionization gas.
For example, [O~I] emission is an important for the cool shell in the CF94
models even though $n\sim 2\times 10^8$ cm$^{-3}$.
If SN~1979C's densities were similar to those of SN~1980K, 
we would predict little [O~II] 3726,3729 \AA \ emission
which has n$_{\rm cr}$ = $5 \times 10^{3}$ cm$^{-3}$, and this agrees
with observation. Indeed, the FOS spectrum of SN~1979C 
reveals strong [O~II] 2470 \AA \ emission but
virtually no [O~II] 3727 \AA \ emission. 

As shown in Table 2, only those emission lines
with critical densities above 10$^{5}$ cm$^{-3}$ are detected in both 79C and 
80K.
Consequently, we conclude that all lines with critical densities 
below 10$^{5}$ cm$^{-3}$ are collisionally suppressed relative to lines having 
higher
critical densities.
Electron densities of $(1-3)\times 10^{6}$ cm$^{-3}$ 
would also explain the observed low ratio of [O~III] (4959 + 5007)/4363.
The [O~III] 4959,5007  \AA \ lines
have n$_{\rm cr}$ = $6.2 \times 10^{5}$ cm$^{-3}$. If the electron densities
were a bit above $1 \times 10^{6}$ cm$^{-3}$ , then the normally
strong [O~III] lines would be suppressed relative to [O~III] 4363 \AA \ 
line intensity (n$_{\rm cr}$ = $2.6 \times$ 10$^{7}$ cm$^{-3}$). 
Furthermore, the [O~III] electron temperature 
must be greater than $\sim$ 15,000 K  
(i.e., I(4959+5007)/I(4363) $\simeq$ 4 assuming n =
$(2-5) \times 10^{6}$ cm$^{-3}$)
in order to generate sufficient 4363 \AA \ emission to be detected given the
S/N of the data. 

A very broad emission feature seen in SN 1979C's 
spectrum near 5800 \AA \ may be partially
due to the temperature sensitive [N~II] 5755 line. If correct,
this would also indicate a [N~II] temperature above 10$^4$ K.
The overall weaker [O~III] emission seen in the SN~1980K spectrum might be due
to low ejecta ionization as indicated by the presence of
of strong [Fe~II]  5050--5400 \AA \ and 7155 \AA \ emission.

\subsection{Observed vs. Predicted Late-Time SNe II-L Emission}

Late-time optical observations of SNe II-L 
can be used to test emission predictions in CF94's recent SN--CSM models.
In Table 1, we list CF94's predicted relative line intensities for 
t = 10 and 17.5 yr arising from SN--CSM interactions assuming 
either a power law or red supergiant (RSG) density gradient.
There is a general agreement with the observed line list, although there
are significant differences with the relative line intensities.
The observed line widths are approximately those expected and
the general picture of emission from circumstellar interaction is
supported, but CF94 present two detailed models that appear not to
be well matched to the conditions in SN 1980K and SN 1979C.
The models are based on the explosion of a $20\Msun$ red supergiant
with two variations for the outer density profile.
Although not ideal, these models do provide a starting point for discussing the 
current
observations.
The RSG model predicts about the right H$\alpha$ luminosity
but weaker [O~I] and stronger [O~III] (relative to H$\alpha$) than observed in
all four cases.  On the other hand, the 
10 and 17.5 yr old power law models predict about the right
H$\alpha$/[O~I] ratio, but less [O~II] 7319,7330  \AA \ emission,
too much [O~III] 4959,5007, and much weaker H$\alpha$ luminosity than observed.

Larger discrepancies between observations and models are seen for several UV 
lines. 
Both RSG and power law models predict Mg~II 2800 \AA \ to be the dominant 
UV/optical
line (after Ly$\alpha$) at a strength $5 - 8$ times more than H$\alpha$. 
However, the SN~1979C spectrum shows Mg~II to be slightly weaker than H$\alpha$, 
with the strongest line actually being C~II] 2324,2325.
Likewise, Mg~I] 4571 \AA \ is predicted to be about half the strength of 
H$\alpha$ 
yet is not seen in either 79C or 80K.
Large differences between observed and predicted Mg II intensities are a 
problem common in young, O-rich SNRs (Blair et al. 1994; Sutherland \& Dopita 
1995) 
and may simply be indicating an incorrect assumed Mg abundance.
However, large differences exist for other UV lines, such as
C~II] 2324,2325 which is about twice the predicted strength,
and this difference grows somewhat larger if internal SN dust extinction is
considered.

Other notable differences involve the [O~II] lines.
We find strong [O~II] 7319,7330 \AA \ emission in both SN~1979C and SN~1980K and 
suspect its
presence in SN~1986E where 
it may have been mis-identified as [Ca~II] 7291,7330 (\cite{CDT95}).
But [O~II] 7319,7330 \AA \ emission is weak in all CF94 models,
due to the high excitation energy of the upper [O~II] level 
and its small collision strength. Because of this CF94 
argued that [Ca~II] was the chief source of line emission observed
near 7300 \AA. However, in view of the new spectra, it now
appears quite likely that the majority
of the 7300 \AA \ emission is due to [O~II]. The presence of strong [O~II] 2470 
\AA \
supports this conclusion. 
Since CF94 do not predict a 2470 \AA \ line strength we cannot
do a direct comparison. But based upon the theoretical [O~II] 2470/7325 ratio of 
0.78,
the model prediction should be about an order of magnitude too low.

Although [O~II] 7319,7330 \AA \ lines are weak in most photoionized nebulae, the
presence of strong [O~II] emission in young SNRs is not uncommon.
In almost all young, O-rich SNRs such as Cas A (SN 1680),
as well as in SN 1957D (Cappellaro et al. 1995) and SN~1986J (Leibundgut et al. 
1991), 
one finds strong [O~II] 7325 \AA \ emission 
which is equal to or stronger than the other forbidden
oxygen line emissions. Strong [O~II] emission has been successfully modeled by 
Sutherland and Dopita (1995) using a mixture of shock and photoionized 
emissions zones in O-rich ejecta.

Additional model-observation differences involve the
the Ca~II IR triplet and the [S III] 9069,9531 \AA \ lines.
These emission lines are predicted to be strong in the power law models but are
not detected in the published near-IR spectrum of SN 1980K (FHM95).
Finally, the favored power law models predict
a steady decline of the H$\alpha$ emission and 
a rapid increase of [O~III] 4959,5007 \AA \ (CF94).

Many model-observation differences appear to be  related to higher electron and 
gas 
densities than the CF94 models assume.
As noted above, the presence of strong [O~III] 4363 \AA \ emission
indicates densities of $10^{6-7}$ cm$^{-3}$. 
High oxygen gas densities would also account for the lack of
any [O~III] 4959 and 5007 line discrimination in the [O ~III] profile
for SN~1979C. At densities above 10$^{6}$  cm$^{-3}$, several
lines predicted to be observable in the models, like [O~II] 3727, 
[N~II] 6548,6583 \AA \ and [S~II] 6716,6731 \AA,  will be strongly de-excited 
(see 
Table 2).

The differences indicate that the CF94 stellar model is not the best
one to model SN 1979C or SN 1980K.
Their assumed circumstellar density may also be a factor.
Models for the light curves of these supernovae suggest a H envelope
mass of $\sim 1\Msun$ (Swartz, Wheeler, \& Harkness 1991;
Blinnikov \& Bartunov 1993; Arnett 1996), much lower than the mass
assumed in CF94.
With the low envelope mass, processed core material may not be
strongly decelerated and could be moving at $\sim 5,000$ km s$^{-1}$.
Some deceleration of the core material could lead to hydrodynamic
instabilities and clumping of the gas, which is consistent with
the spiky profiles seen in the spectral lines of O, C, and Mg.
If the gas is H-depleted core material, strong differences can be
expected with the models of CF94 who assumed cosmic abundances.

The models presented in CF94 also assumed a circumstellar density
determined by a mass loss rate $\dot M=5\times 10^{-5}\ml$ for a wind velocity
of $10\kms$. However, the analysis of the radio turn-on by Lundqvist
\& Fransson (1988) leads to $\dot M=1.2\times 10^{-4}\ml$ for SN 1979C for the
same wind velocity.
There is new evidence for a high circumstellar density around SN 1979C.
Immler, Pietsch, \& Aschenbach (1998) have detected X-ray emission from
the supernova with {\it ROSAT} and determined a $0.1-2.4$
keV X-ray luminosity
of $1.0\times 10^{39}\ergs$.
The summed luminosity of the observed lines listed in Table 1 is
$1.3\times 10^{39}\ergs$ and this figure is likely to be substantially
increased when unobserved lines such as L$\alpha$ are included (CF94).
In the circumstellar interaction model, the optical and ultraviolet
lines are excited by the X-ray radiation, so the implication is that
the X-rays from the reverse shock wave are being absorbed by the cool
circumstellar shell.
This would explain the observed constancy of the line emission because
the X-ray emission at the reverse shock should decline  slowly if
the reverse shock is a cooling shock (CF94).
Application of eq. (2.17) of CF94 shows that the circumstellar shell
can still be optically thick at 1 keV at the time of the observations if
the density power law for the supernova is $n\approx 12$, where
$\rho\propto r^{-n}$, or 
if the circumstellar density is about twice that quoted above.
In the CF94 power law density model at an age of 10 years, the
supernova H density just inside the reverse shock front is 
$4\times 10^4$ cm$^{-3}$.
For SN 1979C, the density may be several times higher, but it still
falls short of the density implied by the [O III] line ratio.
The implication is that the [O III] lines are formed inside of the
reverse shock, which is consistent with the lack of narrowing of
the lines with time.

\section{Conclusions}

From Keck and MDM optical spectra of SN 1980K and optical MMT and UV {\it HST}
spectra of SN 1979C, we find the following:

1) The optical spectrum of SN~1980K taken at 15 and 17 yr  
shows continued strong and broad 5500 km s$^{-1}$ emission lines 
of H$\alpha$, [O~I] 6300,6364 \AA, and [O~II] 7319,7330 \AA,
with weaker but similarly broad lines of [S~II] 4068,4072 \AA,
H$\beta$, [Fe~II] 7155 \AA, and a [Fe~II] blend at 5050--5400 \AA.
The presence of [S~II] 4068,4072 \AA \ but a lack of 
[S~II] 6716,6731 \AA \ emission suggests collisional de-excitation 
of the [S~II] 6716 and 6731 \AA \ lines due to electron densities of 10$^{5-6}$ 
cm$^{-3}$.
The 1997 MDM spectra indicates a H$\alpha$
flux of $1.3 \pm 0.2 \times 10^{-15}$ erg cm$^{-2}$ s$^{-1}$, suggesting
a 25\% drop from 1987--1992 levels sometime during the period 1994 to 1997, 
possibly simultaneously with an observed decrease in nonthermal radio emission.
 
2) Like SN~1980K, SN~1979C's optical spectrum at t = 14.0 yr shows 
$\sim$ 6000 km s$^{-1}$ wide emission lines
but weaker H$\alpha$,
strong [O~III] 4959,5007 \AA, clumpy [O~I] and [O~II] line profiles,
no detectable [Fe~II] 7155 \AA \ emission, and a faint but very broad
emission feature near 5750 \AA. A 1997 {\it HST} Faint Object Spectrograph
spectrum covering the range 2200 -- 4500 \AA \ shows strong lines of 
C~II] 2324,2325 \AA,
[O~II] 2470 \AA, and Mg~II 2796,2803 \AA \ along with weaker
[Ne~III] 3969 \AA, [S~II] 4068,4072 \AA, and [O~III] 4363 \AA.
A lack of [O~II] 3726,3729 \AA \ emission together with
a [O~III] (4959+5007)/4363 $\simeq$ 4 indicates electron densities $\sim$
10$^{6-7}$ cm$^{-3}$. 
Furthermore, the [O~III] temperature must be greater than $\sim$ 15,000 K in
order to generate sufficient 4363 emission to be so easily detected 
(i.e., I(4959+5007)/I(4363) $\simeq$ 4 assuming n = $(2-5) \times 10^{6}$ 
cm$^{-3}$). 
A very broad emission feature seen in SN~1979C's 
spectrum near 5800 \AA \ may be partially
due to the [N~II] line at 5755 \AA.

3) In both SN~1979C and SN~1980K, several lines show one or more sharp emission 
peaks.
The blueward peak(s) are substantially stronger than those toward the red 
indicating
internal dust extinction with the expanding ejecta.
The amount of internal extinction in SN1979C is estimated to be E(B--V) = $0.11 
- 0.16$ mag. 
The line profile differences exhibited between H$\alpha$ and the oxygen lines
suggests that we are seeing emission from two or more separate regions, possibly
the shell (H$\alpha$) and the inner SN ejecta (O lines).

4) Comparison of these observations to late-time SN model predictions
indicates several areas of significant differences, many of which can be
attributed to model electron densities being several orders of magnitude too 
low.
For example, [O~II] 3727 emission
is predicted to be moderately strong in the models but is not seen 
due to densities well above the [O~II]
critical density of $4 \times $ 10$^{3}$ cm$^{-3}$.
In other cases, such as Mg ~II 2800 which arises chiefly from the swept-up 
shell, 
the observed emission is far weaker than predicted.
The parameters for future models will have to be more closely adapted to
the conditions in these supernovae.

It should be noted that other SNe~II besides Type II-L can exhibit bright, 
late-time optical emission, 
most notably the Type~IIn
objects SN~1986J (\cite{Lei91}) and SN~1988Z (\cite{Fil91}; \cite{SS91}; and 
\cite{Turatto93}). 
These objects show quite a different optical spectrum from the SNe~II-L 
discussed above
and are believed to be encountering
dense, clumpy CSM (\cite{Chugai93}; \cite{CD94}).
The current, single Type II-P detection of 
SN~1923A in M83 in the radio (\cite{Eck97}, \cite{Weiler98} ) 
together with the lack of any reported late-time optical detections of
SNe II-P suggests significant mass loss differences between SNe II-L and II-P.
It will be interesting to see if SN~1923A can be recovered optically and how its 
UV and optical
emission properties compare to those of SNe~II-L.

Into what kind of young remnants will these SN II-L
emission nebulae evolve? 
Except for the higher densities and the presence of strong hydrogen emission, 
the observed late-time spectra of SNe II-L bear some resemblance to 
Galactic and LMC O-rich SNRs such as Cas A and
1E0102-7219 (\cite{KC77}; \cite{Blair89}).
Their O, S, and C emission-line dominated spectrum with expansion
velocities around 5000 km s$^{-1}$ is 
not unlike that seen in 10$^{2-3}$ yr old O-rich SNRs.
The weakness of [O~III] 4959,5007 emission and the absence of lines like 
[O~II] 3727 \AA, [Ne~IV] 2425 \AA, and [Ne V] 3426 \AA \ can be 
attributed to higher filament densities and lower ionization levels.

However, it is not at all clear if the progenitors of these older, metal-rich 
remnants are related to SNe II-L events. In the case of Cas A, the
progenitor may have been a WN star, quite different from the RSG
progenitor usually assumed for SNe II-L. Moreover, the Cas A supernova appears 
to have been
subluminous and thus very different from the super-luminous SN II-L events 
SN~1979C and SN~1980K.
This all raises the issue of whether SNe of different types can
yet leave similar looking young remnants at ages 10 to 100 yr.
If SN~1979C and SN~1980K remain luminous 
for several more decades, we may be able to address this question directly.

\acknowledgments

We are grateful to the Keck, MMT, and MDM observatory staffs for their excellent
observing assistance, and K. Weiler and M. Montes for
generously 
communicating their radio study results on SN~1980K prior to its publication. 
Financial support for this work was provided by NASA through Grant Nos. GO-6043
and GO-6584 from the Space Telescope Science Institute, which is operated
by AURA, Inc., under NASA Contract No. NAS 5-26555. We also acknowledge
NSF grants AST-9529232 to R.A.F. and AST-9417213 to A.V.F.

\clearpage
\begin{deluxetable}{lccccccc}
\footnotesize
\tablecaption{Observed and Predicted Late-Time Emission from SNe II-L}
\tablewidth{0pt}
\tablehead{
\colhead{}  & 
\colhead{SN 1986E$^{a}$}  & \colhead{SN 1980K$^{b}$} &
\colhead{SN 1979C$^{c}$} & \colhead{SN 1970G$^{d}$} &
\colhead{RSG} & \colhead{Power Law} & \colhead{Power Law} \\
\colhead{Item ~ ~ ~}     &
\colhead{NGC 4302} & \colhead{NGC 6946} & \colhead{NGC 4321} & \colhead{NGC 
5457} &
\colhead{Model$^{e}$} & \colhead{Model$^{e}$} & \colhead{Model$^{e}$} }
\startdata
Age at Observation (yr)   & 8     & 14-15    & 12--14    & 22    & 10        & 
10      & 17.5    \\
Distance  ~ ~ (Mpc)       &16.8   &  7.5  &   17  &    7  &   \nodata & \nodata 
& \nodata \\
E(B--V) ~ ~ ~(mag)        &  0.02 & 0.40  &  0.23 &  0.15 &   \nodata & \nodata 
& \nodata \\ 
Mass loss rate$^{f}$      &  4.7  & 2.0   &   19  &   2   &   \nodata & \nodata 
& \nodata \\
H$\alpha$ Luminosity$^{g}$&  8.9  & 2.5   &   15  &   1   &   4.8     & 0.9     
& 0.5    \\
Peak Radio Luminosity$^{h}$ & 1.1 & 1.0   &   26  &  1.5  & \nodata   & \nodata 
& \nodata \\
Relative Line Fluxes   &\underline{F($\lambda$)~~I($\lambda$)} 
&\underline{F($\lambda$)~~I($\lambda$)} & \underline{F($\lambda$)~~I($\lambda$)} 
&\underline{F($\lambda$)~~I($\lambda$)} & 
 &   &       \\
 ~ ~ H$\alpha$ 6563     & 10~~~10  & 10~~~10 & 10~~~10  & 10~~~10  &    10     &  
 10    &  10   \\
 ~ ~ [O I] 6300,6364    & ~9~~~~9  & ~7~~~~7 & ~9~~~~9  & ~6~~~~6  &     1     &  
  5    &   7   \\
 ~ ~ [O II] 2470,2470   &\nodata   & \nodata & ~2~~~~6  &\nodata   & \nodata   & 
\nodata &\nodata \\
 ~ ~ [O II] 7319,7330   & ~3~~~~3  & 10~~~~9 & 12~~~11  &\nodata   &  \nodata  &  
  1    &   1   \\
 ~ ~ [O III] 4363       &\nodata   & ~1~~~~2 & ~2~~~~3  &\nodata& \nodata   & 
\nodata &\nodata \\
 ~ ~ [O III] 4959,5007&~$<$3~~$<$5~~~&~2~~~~3&~9~~~11   &~$<$6~~$<$7~~~& 34     
& 
  23    &  46 \\   
~ ~ C II 2324,2325     &\nodata &\nodata& ~6~~~25  &\nodata&    11     &   15    
&  10   \\
  ~ ~ Mg II 2796,2803    &\nodata &\nodata& ~4~~~~9  &\nodata&    49     &   79    
&  57   \\
 ~ ~ [Ne III] 3868,3869 &\nodata   &\nodata  & ~2~~~~3  &\nodata& 3   & 
4 & 5 \\
 ~ ~ [S II] 4068,4076   &\nodata   & ~2~~~~4 & ~1~~~~2  &\nodata&  \nodata  &    
2    &   3   \\
 ~ ~ [Fe II] 7155       &~$<$3~~$<$3~~~&~3~~~~3&~$<$5~~$<$5~~~&\nodata&\nodata & 
\nodata &\nodata \\
                       &     &       &       &       &           &         &     
  \\
Detection Upper Limits &     &       &       &       &           &         &     
  \\
 ~ ~ [O II] 3726,3729   &\nodata &\nodata&$<1$~~$<$2~~~&\nodata& $<$1 & 2   & 5   
  \\
 ~ ~ Mg I] 4571         &\nodata &~$<$2~~$<$4~~~&~$<$2~~$<$4~~~&\nodata& 3  &  6  
&   4  \\
 ~ ~ [Na I] 5890,5896   &~$<$3~~$<$4~~~&~$<$2~~$<$2~~~&?&~$<$5~~$<$6~~~&  2  & 
10 
 &  13   \\
 ~ ~ [N II] 6548,6583   
&~$<$3~~$<$3~~~&~$<$5~~$<$5~~~&~$<$5~~$<$5~~~&~$<$5~~$<$5~~~& 
                                                                    $<$1 &  5  & 
 7   \\
 ~ ~ [C I] ~~8729       &\nodata &~$<$2~~$<$2~~~&\nodata&\nodata&   0       &    
2    &   4  \\
 ~ ~ [S III] 9069,9531  &\nodata &~$<$3~~$<$3~~~&\nodata&\nodata&   2       &   
11    &  13   \\
                       &     &       &       &       &           &         &     
  \\
Expan. Velocity ~(km s$^{-1}$)&  &       &       &       &           &         & 
      \\
 ~ ~ H$\alpha$ 6563 (blue)&$-6000$  &$-5700$&$-6300$ &$-5400$& \nodata   & 
\nodata & \nodata \\
 ~ ~ ~ ~ ~ ~ ~ ~ ~(red)   &$+2600$  &$+5500$&$+6300$ &$+5300$& \nodata   & 
\nodata & \nodata \\
 ~ ~ [O I] 6300,6364 (blue)&$-7350$ &$-6000$&$-6300$ &$-6450$& \nodata   & 
\nodata & \nodata \\
 ~ ~ [O II] 7319,7330 (blue)    &\nodata &$-5000$&$-6300$&\nodata& \nodata   & 
\nodata & \nodata \\
 ~ ~ ~ ~ ~ ~ ~ ~ ~ ~ ~ ~ ~ (red)&\nodata &$+5000$&$ +300$&\nodata& \nodata   & 
\nodata & \nodata \\
 ~ ~ [S II] 4068,4076 (blue)     &\nodata&$-5500$&\nodata&\nodata& \nodata   & 
\nodata & \nodata \\
 ~ ~ ~ ~ ~ ~ ~ ~ ~ ~ ~ ~ ~ (red) &\nodata&$+4000$&\nodata&\nodata& \nodata   & 
\nodata & \nodata \\
H$\alpha$ Line Asymmetry & red?   & blue  &  blue & blue  & \nodata   & \nodata 
& \nodata \\
Clumpy [O I] Emission ? & yes?    & yes   &  yes  &  yes  & \nodata   & \nodata 
& \nodata \\
Narrow nebular lines ? & yes      & yes   &  yes  &  yes  & \nodata   & \nodata 
& \nodata \\
\enddata
\tablenotetext{}{NB: I($\lambda$) are observed line fluxes, F($\lambda$) 
                     fluxes corrected for reddening and narrow
                     line components; \\
~ ~ ~ ~ ~ ~ ~ ~ ~ ~ listed relative to H$\alpha$ = 10.}
\tablenotetext{a}{\cite{CDT95}. Note: We identify [O II] instead of [Ca II] as 
                                the line observed at 7300 \AA \ in SN 1986E.}
\tablenotetext{b}{This paper and \cite{Fesen95}.}
\tablenotetext{c}{This paper and \cite{FM93}.}
\tablenotetext{d}{\cite{Fesen93}.}
\tablenotetext{e}{\cite{CF94}.}
\tablenotetext{f}{In units of 10$^{-5}$ yr$^{-1}$ M$_{\odot}$; values are from 
\cite{Montes97}
                  and \cite{Weiler93}. }
\tablenotetext{g}{In units of 10$^{37}$ erg s$^{-1}$; values calculated using
    E(B--V) and distances quoted.}
\tablenotetext{h}{Peak 6 cm luminosity in units of 10$^{26}$ erg s$^{-1}$ 
Hz$^{-1}$;
    \cite{Montes97}, \cite{Weiler93}, and \cite{Cowan91}.}
\end{deluxetable}

\begin{center}
\newpage
\doublespace
\renewcommand{\arraystretch}{1}
\begin{deluxetable}{llccc}
\footnotesize
\tablecaption{Critical Densities of Detected and Non-Detected Lines}
\tablewidth{0pt}
\tablehead{
\colhead{Density Range/}                     &
\colhead{Critical Density$^{a}$} &
\colhead{Line Strength in}  &
\colhead{Line Strength in} & \colhead{Power Law Model} \\
\colhead{Emission Lines}                     &
\colhead{~ ~ ~ ~(cm$^{-3}$)~ ~ ~ ~} &
\colhead{SN 1980K at 15 yr}  &
\colhead{SN 1979C at 14 yr} & \colhead{Predictions at 17.5 yr} }
\startdata
n$_{e}$ $\leq$ 10$^{5}$ cm$^{-3}$   &                       &         &         
&  \\
 ~[N II] 6548,6583  & $8.0 \times 10^4$                      & absent  & absent  
& strong    \\
 ~[O II] 3726,3729  & $2.6 \times 10^3$, $4.8 \times 10^3$   & absent  & absent  
& moderate   \\
 ~[S II] 6716,6731  & $1.6 \times 10^3$, $1.5 \times 10^{4}$ & absent  & absent  
& weak     \\
 ~[Ne IV] 2422,2425 & $3.8 \times 10^4$, $1.4 \times 10^5$   & \nodata & absent  
& weak     \\
                   &                                        &         &         
&  \\
n$_{e}$ = 10$^{5-6}$ cm$^{-3}$  &                           &         &         
&  \\
 ~[O III] 4959,5007  & $6.2 \times 10^5$                     & weak    & 
moderate& v. strong \\
 ~[S III] 9069,9531 & $8.0 \times 10^5$                      & absent? & \nodata 
& strong     \\
                    &                                       &         &         
&  \\
n$_{e}$ $\geq$ 10$^{6}$ cm$^{-3}$   &                       &         &         
&  \\
 ~[O I] 6300,6364    & $1.6 \times 10^6$                     &  strong &  strong 
& strong     \\
 ~[O II] 2470,2470   & $3.3 \times 10^6$, $5.7 \times 10^6$  & \nodata & strong  
& \nodata     \\
 ~[O II] 7319,7330   & $5.7 \times 10^6$                     & strong  &  strong 
& weak    \\
 ~[O III] 4363       & $2.6 \times 10^7$                     & weak?   & weak    
& \nodata  \\
 ~[S II] 4069,4076   & $1.3 \times 10^6$, $1.7 \times 10^6$  & weak    & weak    
& moderate   \\
 ~[Ne III] 3868,3869 & $8.5 \times 10^6$                     & \nodata & 
moderate& moderate     \\
 ~[Ne V] 3346,3426  & $1.4 \times 10^7$                      & \nodata & absent  
& moderate   \\
 ~~C II] 2324,2325   & $3.0 \times 10^8$, $7.2 \times 10^9$  & \nodata & strong  
& strong     \\
\enddata
\tablenotetext{a}{Evaluated for T= 10$^4$ K; values taken from IRAF/STSDAS task 
nebular.ionic
                  by D. Shaw.}
 
\end{deluxetable}
\end{center}

\clearpage

\clearpage

\begin{figure} 
\plotone{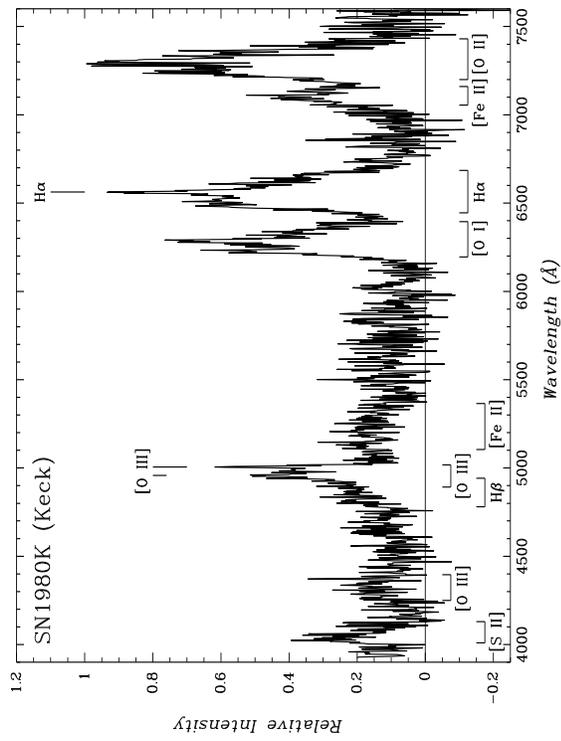}
\figcaption[fesenr.fig1.ps]{Keck spectrum of SN 1980k taken August 1995. 
  \label{fig1}}
\end{figure}

\begin{figure}
\plotone{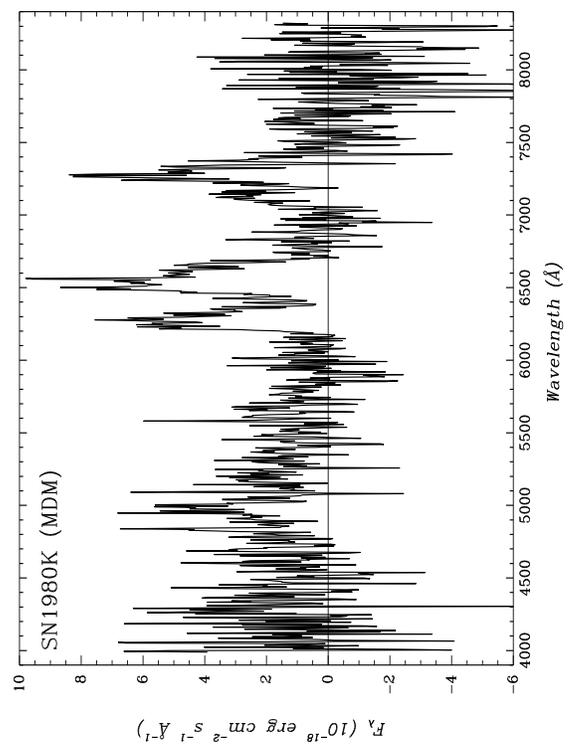}
\figcaption[fesenr.fig2.ps]{MDM spectrum of SN 1980K taken November 1997.
  \label{fig2}}
\end{figure}

\begin{figure}
\plotone{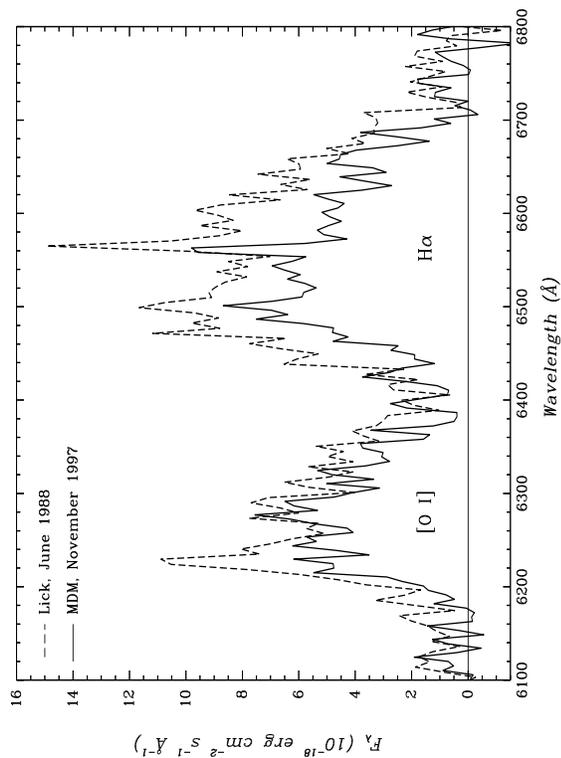}
\figcaption[fesenr.fig3.ps]{Comparison of June 1988 and November 1997 SN 1980K
   spectra for the [O~I] $\lambda\lambda$6300,6364 and H$\alpha$ emission 
profiles.
  \label{fig3}}
\end{figure}

\begin{figure}
\plottwo{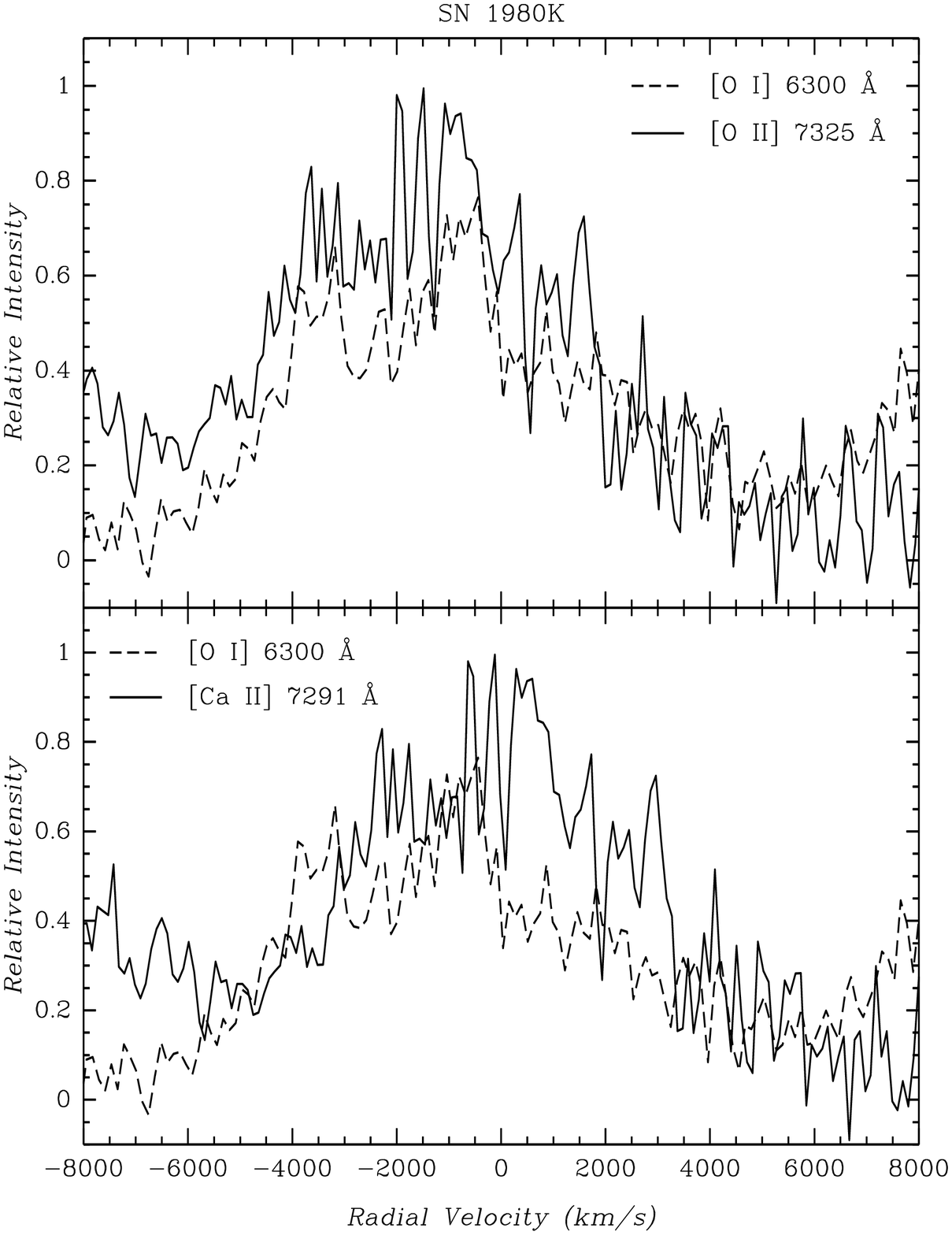} {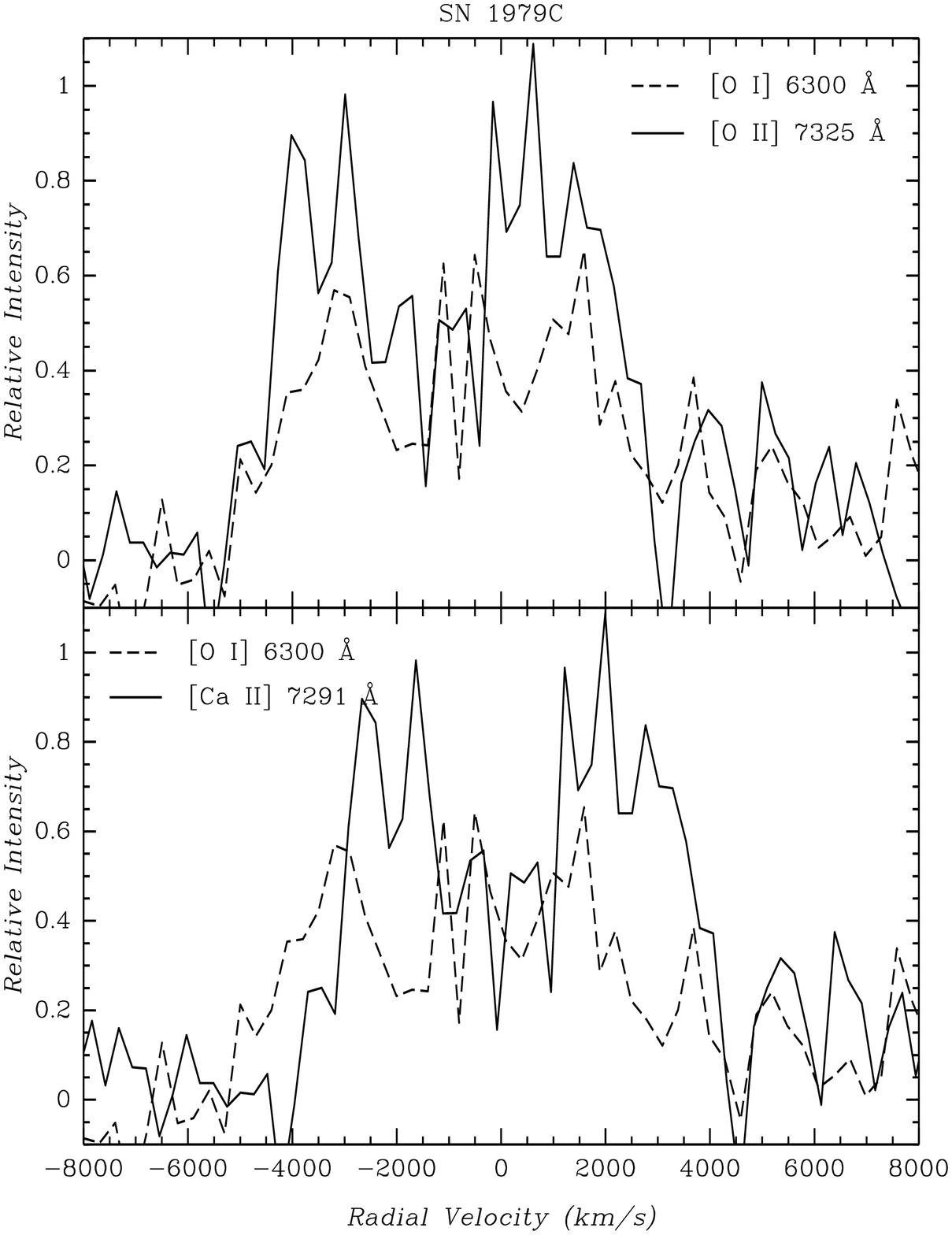}
\caption{Line profile comparison of SN~1980K's (Fig. 4a) and
   SN~1979C's (Fig. 4b) [O~I] $\lambda\lambda$6300,6364
   emission feature  vs. the 7300 \AA \ emission feature in velocity space when 
interpreted as
   due to [O~II] $\lambda\lambda$7319,7330 (top) or [Ca~II] 
   $\lambda\lambda$7291,7324
   (bottom) Velocity scales given are calculated for the [O~I] 6300 \AA \ line,
    the [Ca~II] 7291 \AA \ line, or the [O~II] line blend at 7325 \AA.
  \label{fig4}}
\end{figure}

\begin{figure}
\plotone{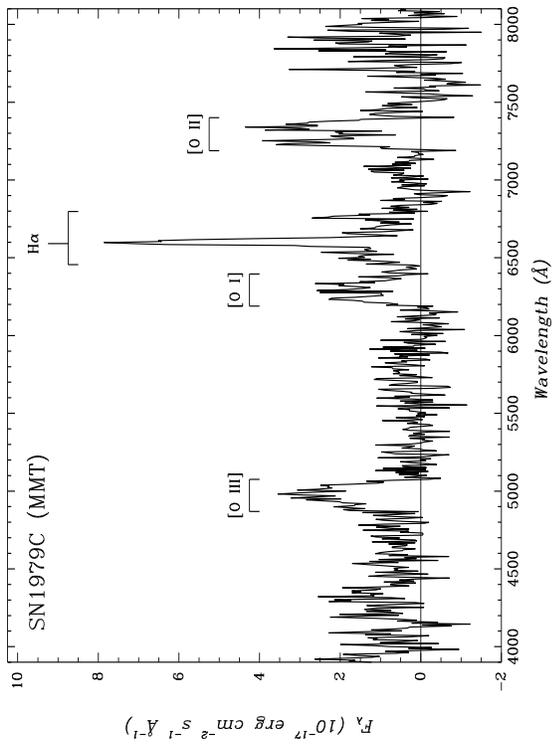}
\figcaption[fesenr.fig5.ps]{MMT spectrum of SN 1979C taken May 1993.
  \label{fig5}}
\end{figure} 

\begin{figure} 
\plotone{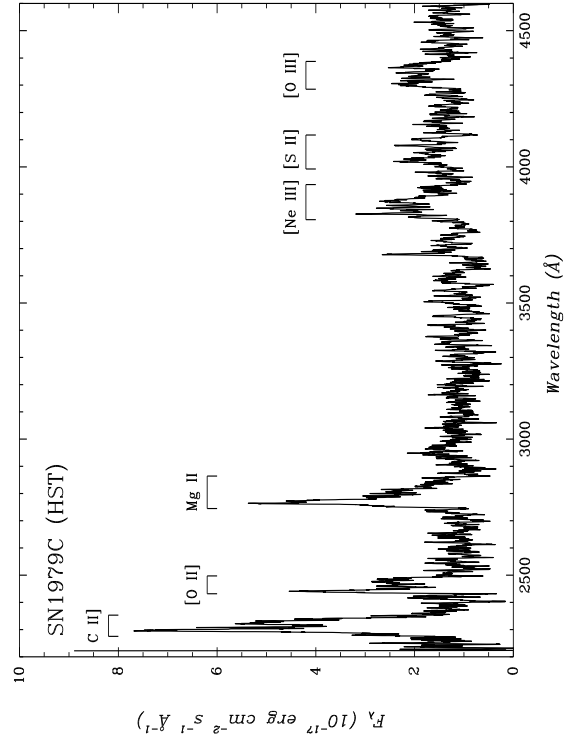}
\figcaption[fesenr.fig6.ps]{1997 FOS spectrum of SN~1979C.
  \label{fig6}}
\end{figure}

\begin{figure}
\plotone{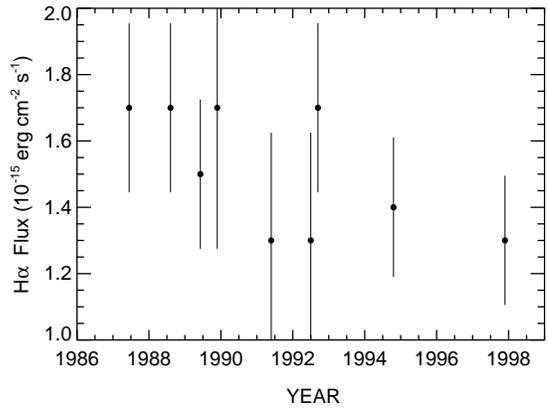}
\figcaption[fesenr.fig7.ps]{Plot of observed H$\alpha$ flux measurements
  for SN~1980K taken since 1987.
  \label{fig7}}
\end{figure}
 
\end{document}